\documentclass[a4paper,12pt]{article}
\usepackage{amsmath}
\usepackage{amssymb}

\numberwithin{equation}{section}

\def\A{\mathcal A}
\def\B{\mathcal B}
\def\C{\mathcal C}
\def\NN{\mathbb N}
\def\RR{\mathbb R}
\def\ZZ{\mathbb Z}

\newcommand{\suppress}[1]{}

\begin{document}
\newtheorem{theo}{Theorem}
\newtheorem{prop}{Proposition}
\newtheorem{lem}{Lemma}
\newtheorem{cor}{Corollary}
\newtheorem{defi}{Definition}

\title{On q-ary Codes Correcting All Unidirectional Errors of a
Limited Magnitude }
\date{}
\author{R.\ Ahlswede, H.\ Aydinian, L.H.\ Khachatrian
and L.M.G.M \ Tolhuizen
\thanks{L.M.G.M.\ Tolhuizen is with Philips Research Laboratories, Prof.\ Holstlaan 4,
5656 AA Eindhoven,
   The Netherlands;
\sf{ludo.tolhuizen@philips.com}}}
\maketitle

\centerline{Dedicated to the memory of Rom Varshamov}

\vskip 1.5truecm

\begin{abstract}

We consider codes over the alphabet $Q=\{0,1,\ldots ,q-1\}$
intended for the control of unidirectional errors of level $\ell$.
That is, the transmission channel is such
that the received word cannot contain both a component larger
than the transmitted one and a component smaller than the
transmitted one.
Moreover, the absolute value of the difference between a transmitted
component and its received version is at most $\ell$.\\ \\
We introduce and study $q$-ary codes capable of correcting all
unidirectional errors of level $\ell$.
Lower and upper bounds for the maximal size of those codes are
presented.\\

We also study codes for this aim that are
defined by a single equation on the codeword coordinates
(similar to the Varshamov-Tenengolts codes for correcting binary
asymmetric errors).
We finally consider the problem of detecting all unidirectional
errors of level $\ell$.
\end{abstract}

\vfill

\eject

\section{Introduction}

An extensive theory of error control coding has been developed
(cf.\ \cite{Hand},\cite{McWSl},\cite{LiCo})
under the assumption of symmetric errors in the data bits;
i.e. errors of type $0\to1$ and $1\to0$ can occur simultaneously in a
codeword.

However in many digital systems such as fiber optical communications
and optical disks the ratio between probability of errors of type
$1\to0$ and $0\to1$ can be large.
Practically we can assume that only one type of errors
can occur in those systems. These errors are called asymmetric.
Thus the binary asymmetric channel, also called $Z$-channel
(shown in Figure.~1),

\begin{center}
\setlength{\unitlength}{0.5 cm}

\begin{picture}(12,6)
\put(1,1){0}
\put(9,1){0}
\put(1,4.5){1}
\put(9,4.5){1}
\put(1.5,1.1){\line(1,0){7}}
\put(1.5,4.6){\line(1,0){7}}
\put(1.5,1.1){\line(2,1){7}}
\end{picture}

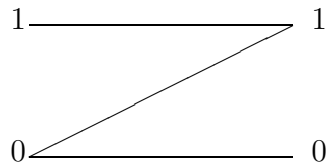
Figure~1: the Z-channel
\end{center}

has the property that a transmitted 1 is always received correctly
but a transmitted 0 may be received as a 0 or 1.

Unidirectional errors slightly differ from
asymmetric type of errors: both $1\to0$ and $0\to1$ type of errors are
possible, but in any particular word all the errors are of the same
type.
The statistics  shows that in some of
LSI/VLSI ROM and RAM memories the most likely faults are of the
unidirectional type.
The problem of protection against unidirectional errors arises also in
designing of fault-tolerant sequential machines, in write-once memory
system, in asynchronous systems etc.

Clearly any code capable of correcting (detecting) $t$-symmetric
errors can be also used to correct (to detect) $t$-unidirectional or
$t$-asymmetric errors. Obviously also any $t$-unidirectional error
correcting (detecting) code is capable of correcting (detecting)
$t$-asymmetric errors. Note that there are $t$-asymmetric error
correcting codes with higher information rate than that of
$t$-symmetric error correcting codes
(\cite{Va73},\cite{CoRa},\cite{HeKl}).
For constructions of codes correcting unidirectional errors see
\cite{We89} and \cite{FaTi}.
It can be shown that the detection problems for
asymmetric and unidirectional errors are equivalent
(see \cite{Bor})
i.e. any $t$-error detecting asymmetric code
is also a $t$-error detecting unidirectional code.

First results on asymmetric error correcting codes are due to
Kim and Freiman \cite{KF}, and Varshamov \cite{Var64},\cite{Va64}.
In \cite{Var64}  Varshamov introduced a metric for asymmetric
errors and obtained bounds for codes correcting
asymmetric errors.
In \cite{Va64}   Varshamov ( and later Weber et al. \cite{We89})
proved that linear codes capable of correcting $t$-asymmetric
errors are also capable of correcting $t$-symmetric errors.
Thus only non-linear constructions
may go beyond symmetric error correcting codes.

In 1965  Varshamov and Tennengolts gave the first construction of
nonlinear codes correcting asymmetric errors \cite{VaTe}.

The idea behind these codes (which we call VT-codes) is surprisingly simple.
Given $n\in\NN$ and an integer $a$ the VT-code $\C(n,a)$
is defined by
\begin{equation}\label{Eq:1.1}
\C(n,a)=\left\{(x_1,\dots,x_n)\in\{0,1\}^n:
\sum_{i=1}^n ix_i\equiv a \;\; (\mod m)\right\}
\end{equation}
where $m\geq n+1$ is an integer.

Varshamov and Tennengolts showed that the code $\C(n,a)$ is capable of
correcting any single asymmetric error. Moreover taking $m=n+1$ there
exists an $a\in\{0,\dots,n\}$ so that
\begin{equation}\label{Eq:1.2}
|\C(n,a)|\geq\frac{2^n}{n+1}.
\end{equation}

Recall that for the maximum size of binary single symmetric error
correcting codes we have
\begin{equation}\label{Eq:1.3}
A(n,1)\leq\frac{2^n}{n+1}.
\end{equation}

Varshamov \cite{Va65} showed that $|\C(n,0)|\geq|\C(n,a)|$.

A number theoretical result due to
von Sterneck (1902) \cite[p.\ 87]{Di} allows to determine the weight
distribution of VT-codes.
This result and its special cases were rediscovered
many times (see \cite{Gin},\cite{Ma},\cite{Maz},\cite{Stan}).
From a practical point of view VT-codes have the advantage of a very
simple decoding algorithm. For systematic encoding of VT-codes see
\cite{AbGaFe} and \cite{BoAlBa}.

In general we call a code of length $n$, correcting $t$-asymmetric
errors a VT-code if it is given by the set of solutions
$(x_1,\dots,x_n)\in\{0,1\}^n$ of a congruence (or several congruences)
of the type
\begin{equation}\label{Eq:1.4}
\sum_{i=1}^n f(i)x_i\equiv a(\mod M)
\end{equation}
where $f:[n]\to\ZZ$ is an injection, $a$ and $M$ are integers.\\
We note that there are deep relationships between VT-codes and
some difficult problems in Additive Number Theory
\cite{Va73}, \cite{Er}.

The idea of VT-codes was further developed by Constantin and Rao
\cite{CoRa},
(see also Helleseth and Kl\o ve \cite{HeKl}) by constructing
group-theoretical codes based on Abelian Groups.\\
Levenshtein noticed that VT-codes can also be used to correct single
insertion/deletion errors \cite{Le}.

Modifications of VT-codes where used to construct new codes
correcting $t$-asymmetric errors
\cite{Va73}, \cite{Na}, \cite{GeMh}, \cite{BoCu}
and bursts of errors \cite{OgYa}, \cite{VaZo}
(see also \cite{BoAlBa}, \cite{DePi}, \cite{FaTi} for other
constructions).
For an excellent survey on the results in this direction
see Kl\o ve \cite{Kl2}. \\
Very few constructions are known for codes correcting unidirectional
errors (for more information see \cite{Bl}). Note that VT-codes (1.1)
and its known modifications are not capable
of correcting unidirectional errors.

In 1973 Varshamov introduced a $q$-ary asymmetric channel
\cite{Va73}.

The inputs and outputs of the channel are $n$-sequences over
 the  $q$-ary alphabet  $Q=\{0,1,\dots,q-1\}$.
If the symbol $i$ is transmitted then the only symbols which the
receiver can get are $\{i,i+1,\dots,q-1\}$.
Thus for any transmitted vector $(x_1,\dots,x_n)$ the received vector
is of the form $(x_1+e_1,\dots,x_n+e_n)$ where $e_i\in Q$
and
\begin{equation}\label{Eq:1.5}
x_i+e_i\leq q-1,\ i=1,\dots,n.
\end{equation}

Then it is said that $t$-errors have occurred if
$e_1+\dots+e_n=t$.
Generalizing the idea of VT-codes, Varshamov \cite{Va73}
presented  several constructions of $t$-error correcting codes
for the defined channel.
These codes have been shown in \cite{McE} to have larger cardinality
than BCH codes correcting  $t$ errors for $q\geq2$ and for large $n$.

We continue here the work started in \cite{ahlsayd}.
We consider
{\bf a special type of asymmetric errors in a $q$-ary channel},
where the magnitude of each component of ${\bf e}$ satisfies
$0\leq e_i\leq\ell$ for  $i=1,\dots,n$.
We refer to $\ell$ as level.

Correspondingly we say that
{\bf an unidirectional error of level $\ell$ has occurred},
if the output is either {\bf x} + {\bf e} or {\bf x} - {\bf e}
(in the latter case, it is of course required
that $x_i\geq e_i$ for all $i$).

If the error vector {\bf e} has Hamming weight $d_H({\bf e})=t$, then
we say that $t$ errors of level $\ell$ have occured.

Thus the general problem is the following.

Given $n,\ell,t,q$ construct $q$-ary codes of length $n$ capable of
correcting $t$ errors of level $\ell$. Of course we wish the size of a
code to be as big as possible.

Note the difference between the channel described above and
Varshamov's channel when $q>2$. This is shown for $q=3$, $l=1$,
$t\geq 2$ in Figure~2.

\vskip 1truecm

\setlength{\unitlength}{0.5 cm}

\begin{picture}(22,8)
\put(1,1){0}
\put(9,1){0}
\put(1,4.5){1}
\put(9,4.5){1}
\put(1,8){2}
\put(9,8){2}
\put(1.5,1.1){\line(1,0){7}}
\put(1.5,4.6){\line(1,0){7}}
\put(1.5,8.1){\line(1,0){7}}
\put(1.5,4.6){\line(2,1){7}}
\put(1.5,1.1){\line(2,1){7}}

\put(13,1){0}
\put(21,1){0}
\put(13,4.5){1}
\put(21,4.5){1}
\put(13,8){2}
\put(21,8){2}
\put(13.5,1.1){\line(1,0){7}}
\put(13.5,4.6){\line(1,0){7}}
\put(13.5,8.1){\line(1,0){7}}
\put(13.5,4.6){\line(2,1){7}}
\put(13.5,1.1){\line(2,1){7}}
\put(13.5,1.1){\line(1,1){7}}
\end{picture}

\hskip 5truecm Figure 2:

Asymmetric errors with level 1 \qquad
\ Varshamov's channel \\

In this paper we consider $q$-ary codes correcting all
asymmetric errors of given level $\ell$, (that is $t=n$)
for which we use the abbreviation $\ell$-AEC code, and
$\ell$-UEC codes that correct all unidirectional errors of level
$\ell$.
As above our alphabet is $Q\triangleq\{0,1,\dots,q-1\}$.

In Section~\ref{Sec:dist} we define distances that capture the
capabilities of a code to correct all asymmetric or unidirectional
errors of level $\ell$.

For given $\ell$, let $A_a(n,\ell)_q$ and $A_u(n,\ell)_q$
denote the maximum number of words in a q-ary AEC code,
or UEC code respectively, of length $n$.
Clearly $A_u(n,\ell)_q\leq A_a(n,\ell)_q$.

In Section~\ref{Sec:AEC} we determine $A_a(n,\ell)_q$ exactly for
all $n,\ell$ and $q$.

In Section~\ref{Sec:UEC} we give upper and lower bounds
on $A_u(n,\ell)_q$, which imply that for fixed $q$ and $\ell$
the asymptotic growth rate for $A_u(n,\ell)_q$ equals that of
$A_a(n,\ell)$.

In Section~\ref{Sec:VT} we study $\ell$-AEC and $\ell$-UEC
codes of VT-type.
It is shown that any $\ell$-AEC code of VT-type can be
transformed into an $\ell$-UEC code of VT-type of equal length
and cardinality.
Upper and lower bounds on the maximum number of codewords in a $q$-ary
$\ell$-UEC code of length $n$ of VT-type are derived.
For certain pairs ($\ell,q$) we give a construction of optimal
$\ell$-UEC codes.

In Section~\ref{Sec:UED} we consider the problem of detecting all
errors of level $\ell$.

\section{Distances and error-correcting capabilities}\label{Sec:dist}

In this section we introduce two distances that capture the
capabilities of a code for correcting all symmetrical and
unidirectional errors of a certain level.
Throughout this section we write $L$ for $[0,\ell]$ (where for
integers $a<b$ we use the abbreviation
$[a,b]\triangleq\{a,a+1,\dots,b\}$).

\begin{defi}\label{def:dist}
For {\bf x} = $(x_1,x_2,\ldots ,x_n)\in Q^n$ and
{\bf y} = $(y_1,y_2,\ldots ,y_n)\in Q^n$,
\begin{eqnarray*}
d_{\max}({\bf x},{\bf y})&=&\max\{|x_i-y_i|:i=1,2,\ldots ,n\}\\
d_u({\bf x},{\bf y})     &=&\left\{\begin{array}{ll}
d_{\max}({\bf x},{\bf y})&\mbox{ if }{\bf x}\geq{\bf y}
                          \mbox{ or }{\bf y}\geq{\bf x},\\
2d_{\max}({\bf x},{\bf y})&\mbox{ if }{\bf x}
                           \mbox{ and }{\bf y}
                           \mbox{ are incomparable},
\end{array} \right.
\end{eqnarray*}
where ${\bf x}\geq {\bf y}$  means that $x_i\geq y_i$ for all $i$.
\end{defi}

Later on for short we will write $d({\bf x},{\bf y})$ for
$d_{\max}({\bf x},{\bf y})$.

Note that $d_u$ does not define a metric:
take {\bf x}=(0,2), {\bf y}=(1,0) and {\bf z}=(1,2).
Then $d_u({\bf x},{\bf y})=4>1+2=d_u({\bf x},{\bf z})+
d_u({\bf z},{\bf y})$.

\begin{lem}\label{daprop}
Let ${\bf x},{\bf y}\in Q^n$.
The two following assertions are equivalent: \\
$(i)$ $d({\bf x},{\bf y})\leq \ell$ \\
$(ii)$ there exist ${\bf e}\in L^n$, ${\bf f}\in L^n$
such that ${\bf x}+{\bf e}={\bf y}+{\bf f}\in Q^n$.
\end{lem}

{\bf Proof.}
Suppose that $(i)$ holds. We define ${\bf e}$ and ${\bf f}$ as
\[ e_i = \max(0,y_i-x_i) \mbox{ and } f_i = \max(0,x_i-y_i),\;\;\;
  i=1,2,\ldots ,n . \]
As $d({\bf x},{\bf y})\leq \ell$, the vectors {\bf e} and {\bf f}
are in $L^n$, and for each $i$,
we have that $x_i+e_i=y_i+f_i=\max(x_i,y_i)\in Q$.
That is $(ii)$ holds. \\
Conversely, suppose that $(ii)$ holds, then for each $i$
we have that  $|x_i-y_i|=|f_i-e_i|\leq\max(f_i,e_i)\leq\ell$,
where the first inequality holds since $e_i$ and $f_i$ both
are non-negative.
\hfill $\Box$ \\ \\
The following proposition readily follows from Lemma~\ref{daprop}.

\begin{prop}\label{prop:da}
A code $\C\subset Q^n$ is an $\ell$-AEC code if and only if
$d({\bf x},{\bf y})\geq\ell+1$ for all distinct {\bf x},{\bf y} in $\C$.
\end{prop}

Note that Proposition 1 and the definition of
$d({\bf x},{\bf y})$ imply that for $\ell\geq q-1$,
an $\ell$-AEC code (and therefore also an $\ell$-UEC code)
contains at most a single codeword.
{\bf For this reason, we assume in the remainder of the paper that
$\ell\leq q-2$.}

\begin{lem}\label{duprop}
Let ${\bf x},{\bf y}\in Q^n$.
The two following assertions are equivalent.\\
$(i)$ ${\bf y}\geq {\bf x}$ and $d({\bf x},{\bf y})\leq 2\ell$, \\
$(ii)$ there exist ${\bf e}\in L^n, {\bf f}\in L^n$ such that
${\bf x} + {\bf e}={\bf y} - {\bf f} \in Q^n$.
\end{lem}

{\bf Proof.}
Suppose that $(i)$ holds. We define {\bf e} and {\bf f} as
\[ e_i = \lceil\frac{1}{2}(y_i-x_i)\rceil \mbox{ and }
   f_i = \lfloor\frac{1}{2}(y_i-x_i)\rfloor, \;\; i=1,2,\ldots ,n. \]
As ${\bf y}\geq {\bf x}$, both {\bf e} and {\bf f} have only
non-negative components and for each $i$, we have that
$f_i\leq e_i\leq\lceil\frac{1}{2}(2l)\rceil=\ell$;
moreover, we obviously have that ${\bf e}+{\bf f}={\bf y}-{\bf x}$.
Finally, for each $i$ we have
that $x_i+e_i=y_i-f_i\leq y_i\leq q-1$,
so ${\bf x}+{\bf e} = {\bf y} - {\bf f}\in Q^n$.
We conclude that $(ii)$ holds. \\
Conversely suppose that $(ii)$ holds.
Then ${\bf y}-{\bf x} = {\bf e} + {\bf f}$
and so ${\bf y}\geq {\bf x}$, and for each
$i$ we have that $|y_i-x_i|=y_i-x_i=e_i+f_i\leq \ell+\ell=2\ell$.
That is $(i)$ holds.
\hfill $\Box$ \\ \\
Combination of Lemma~\ref{daprop} and Lemma~\ref{duprop}
yields the following

\begin{prop}\label{prop:du}
A code $\C\subset Q^n$ is an $\ell$-UEC code if and only if
$d_u({\bf x},{\bf y})\geq 2\ell + 1$ for all distinct
{\bf x}, {\bf y} in $\C$.
\end{prop}

\section{$\ell$-AEC codes}\label{Sec:AEC}

It turns out that $A_a(n,\ell)_q$ can be determined exactly
for all integers $n$ and each $\ell\in Q$.

\begin{theo}\label{Th:AEC}
For all integers $n$ and each $\ell\in Q$,
$A_a(n,\ell)_q=\left\lceil\frac q{\ell+1}\right\rceil^n. $
\end{theo}

{\bf Proof.}
Let $\C\subset Q^n$ be an $\ell$-AEC-code.
Let  $\varphi:Q\to\left\{0,1,\dots
\left\lfloor\frac {q-1}{\ell+1}\right\rfloor\right\}$,
be defined as
$$
\varphi(j)=\left\lfloor\frac j{\ell+1}\right\rfloor,\
j=0,\dots,q-1.$$

For any codeword ${\bf x}=(x_1,\dots,x_n)\in\C$ define
$\varphi^n({\bf x})=\bigl(\varphi(x_1),\dots,\varphi(x_n)\bigr)$.
Clearly $\varphi^n$ is injective:
if ${\bf x},{\bf y}\in\C$ are such that
$\varphi^n({\bf x})=\varphi^n({\bf y})$,
then $|x_i-y_i|\leq\ell$, $(i=1,\dots,n)$, that is,
$d({\bf x},{\bf y})\leq\ell$ and so ${\bf x}={\bf y}$.
This implies that $|\varphi^n(\C)|=|\C|$ and since
$\left\lfloor\frac{q-1}{\ell+1}\right\rfloor+1=
\left\lceil\frac q{\ell+1}\right\rceil$ we get
\begin{equation}\label{AAECup}
|\C|\leq\left\lceil\frac q{\ell+1}\right\rceil^n.
\end{equation}

The code $\C$ defined as
\[\C=\bigl\{(x_1,x_2,\ldots ,x_n)\in Q^n:x_i\equiv 0\mod(\ell+1)
\mbox{ for } i=1,2,\ldots ,n\bigr\} \]
obviously is an $\ell$-AEC code that achieves equality in
(\ref{AAECup}).
A received vector can be decoded by component-wise rounding downwards
to the nearest multiple of $\ell$+1.
\hfill$\square$

\section{$\ell$-UEC codes}\label{Sec:UEC}

In this section, we study $A_u(n,\ell)_q$, the maximum number of words
in a $q$-ary $\ell$-UEC code of length $n$.
As any $\ell$-UEC code is an $\ell$-AEC code, Theorem~\ref{Th:AEC}
implies that
\begin{equation}\label{AUup}
A_u(n,\ell)_q\leq A_a(n,\ell)_q=
\left\lceil\frac{q}{\ell+1}\right\rceil^n.
\end{equation}

In some special cases the upper bound (\ref{AUup}) is met with equality.

\begin{prop}
For all $n$ and $\ell$, $A_u(n,\ell)_{2\ell + 2} = 2^n$.
\end{prop}

{\bf Proof.}
By Proposition 2 the code $\{0,2\ell+1\}^n$ meeting $2^n$ has the
desired property and $A_u(n,\ell)_{2\ell+2}\leq2^n$ by (4.1).
\hfill $\Box$ \\ \\
In Section~\ref{Sec:VT} we will construct $q$-ary $\ell$-UEC
codes of VT type.
For various classes of pairs ($q,\ell$),
(for example, if $\ell+1$ divides $q$), these codes have cardinality
$\lceil\frac{q}{\ell+1}\rceil^{n-1}$ and thus they are
below the upperbound (4.1) only by a multiplicative factor.

We continue the present section with two constructions
for $q$-ary $\ell$-UEC codes valid for all pairs $(q,\ell$).
We denote by $Q_{\ell+1}$ all integers in
$Q=[0,q-1]$ that are multiples of $\ell+1$, that is
\begin{equation}\label{AUup}
Q_{\ell + 1} = \{ m \in \{0,1,\ldots ,q-1\}:
                     m \equiv 0\;(\bmod \; \ell+1) \} =
 \{ a(\ell+1): 0\leq a\leq b-1\} ,
\end{equation}
where
$$
b= |Q_{\ell +1}| = \left\lceil\frac{q}{\ell+1}\right\rceil.$$

It is clear that $d({\bf x},{\bf y})\geq \ell+1$ for any two distinct
words ${\bf x},{\bf y}$ in $Q_{\ell+1}^n$.
In the subsequent two subsections we use $Q_{\ell+1}^n$ to
construct a code with minimum asymmetric distance $\ell$+1
for which any two codewords are incomparable.
Thus we have created a code with undirectional distance
at least $2\ell+2$.

\subsection{Construction 1: taking a subset of $Q_{\ell+1}^n$}

For each $j$ let
\[ C(j) = \{ (x_1,x_2,\ldots ,x_n)\in Q_{\ell+1}^n:
  \sum_{i=1}^n \frac{x_i}{\ell + 1}= j \} . \]
Any two distinct words from $C(j)$ clearly are incomparable
and so $C(j)$ is an $\ell$-UEC code.
It is clear that
\[ |C(j)| = | \{ (y_1,y_2,\ldots ,y_n) \in
      \{ 0,1,\ldots ,b-1\}^n:\sum_{i=1}^n y_i =j \} | . \]
It is known \cite[Thm.\ 4.1.1]{An} that $|C(j)|$ is maximized for
$j=j^{\ast}\triangleq\lfloor\frac{1}{2}n(b-1)\rfloor$.
Moreover, according to \cite[Thm.\ 4.3.6]{An},
the following bounds are valid.

\begin{prop}\label{prop:sqrt}
There exist positive constants $c_1$ and $c_2$ $($depending on
$b=\lceil\frac{q}{\ell +1}
\rceil)$ such that
\[ c_1 \frac{1}{\sqrt{n}}b^n \leq |C(j^{\ast})|
\leq c_2 \frac{1}{\sqrt{n}}b^n . \]
\end{prop}

Proposition~\ref{prop:sqrt} implies the following theorem.

\begin{theo}\label{th:sqrt}
For each integer $q$ and $\ell\in Q$, there is a constant $c>0$ such that
for each $n$,
\[ A_u(n,\ell)_q \geq c \frac{1}{\sqrt{n}}
\lceil\frac{q}{\ell+1}\rceil^n \; . \]
\end{theo}

Clearly, (\ref{AUup}) and Theorem~\ref{th:sqrt} imply that for fixed
$q$ and $\ell$ the asymptotic growth rate of $A_u(n,\ell)_q$ is known.

\begin{cor}\label{Cor1}
For each $q$ and each $\ell\in[0,q-1]$\ \
$\lim_{n\rightarrow\infty} \sqrt[n]{A_u(n,\ell)_q} =
\lceil\frac{q}{\ell+1}\rceil$.
\end{cor}

\subsection{Construction 2: adding tails to words from $Q_{\ell+1}^n$}

In order to formulate our second construction clearly, we cast it in
the form of a proposition. Later we take appropriate values for certain
parameters in this construction to obtain a lower bound on
$A_u(n,\ell)_q$.

\begin{prop}\label{prop:tails}
Let $X\subset Q^n$ be a $\ell$-AEC code.
For ${\bf x}\in X$, let $S({\bf x})$ denote the sum of its entries,
and let $s_1,s_2$ be such that for each
${\bf x}\in X$, $s_1\leq S(x)\leq s_2$.
Let $\phi: [s_1,s_2] \rightarrow Q^m$ be such that for all
$a,b\in [s_1,s_2]$ with $a>b$, there is an $i\in\{1,2,\ldots ,m\}$
such that ($\phi(a))_i < (\phi(b))_i$.
Then $\C= \{({\bf x},\phi(S(x)):{\bf x}\in X\} \subset Q^{n+m}$
is an $\ell$-UEC code.
\end{prop}

{\bf Proof.}
Let ${\bf u}=({\bf x},\phi(S({\bf x})))$ and
${\bf v}=({\bf y},\phi(S({\bf y})))$
be two distinct words in $\C$.
As $d({\bf x},{\bf y})\geq \ell +1$, all we have to show is that
${\bf u}$ and ${\bf v}$ are incomparable.
This is clear if ${\bf x}$ and ${\bf y}$ are incomparable.
Now suppose that ${\bf x}$ and ${\bf y}$ are comparable,
say ${\bf x}\geq {\bf y}$.
Then $S({\bf x}) > S({\bf y})$ and hence, by the
property imposed on $\phi$, $u_j < v_j$ for some $j\in [n+1,n+m]$.
\hfill $\Box$ \\ \\
We now apply the construction from Proposition~\ref{prop:tails}.
Given $s_1$ and $s_2$, we take
$m\triangleq\lceil\log_q(s_2-s_1+1)\rceil$, and define
$\phi(s)$ as the $m$-symbols $q$-ary representation of $s_2-s$.
We choose for $X$ a large subset of $Q_{\ell+1}^n$ such that
$s_2-s_1+1$ is small, so that $m$ can be small.
As shown below we can invoke Chebyshev's inequality
to show the existence of a set $X$ such that
$|X|>\frac{3}{4}b^n$, while $s_2-s_1+1 < K_1\sqrt{n}$
for some constant $K_1$.
As a consequence, $m$ can be as small as
$\frac{1}{2}\log_q n+K_2$ for some constant $K_2$.

\begin{theo}\label{Ludobound}
For each $q$ and $\ell$, there exists a positive constant $K$
such that for each $n$,
\[ A_u(n,\ell)_q \geq K b^n n^{-\frac{1}{2}\log_q b}, \mbox{ where }
     b = \lceil\frac{q}{\ell + 1}\rceil \; . \]
\end{theo}

{\bf Proof.}
We start with the well-known Chebyshev inequality.

\begin{prop}\label{Chebyshev}
Let $Y_1,Y_2,\ldots ,Y_n$ be independent, identically distributed
random variables, each with average $\mu$ and variance $\sigma^2$.
For each $\epsilon > 0$, we have that
\[ \mbox{prob}( |\sum_{i=1}^n Y_i \;\; - n\mu| > \epsilon\cdot n )
      \leq \frac{\sigma^2}{n\epsilon^2} \;\; . \]
\end{prop}

We choose now
$\epsilon=\frac{2\sigma}{\sqrt{n}}$ and get
\begin{equation}\label{eq:app1}
 \mbox{Prob} (|\sum_{i=1}^n Y_i-n\mu | \leq 2\sigma\sqrt{n}) \geq
   \frac{3}{4}.
\end{equation}

In the above, we take each $Y_i$ uniformly distributed in
$Q_{\ell+1}=\{a(\ell+1):0\leq a\leq b-1\}$.
It follows from (4.3) that the set $X$ defined as
\[ X= \{ x\in Q_{\ell+1}^n:n\mu-2\sigma\sqrt{n} \leq
   \sum_{i=1}^n x_i \leq n\mu + 2\sigma\sqrt{n} \} \]
has cardinality at least $\frac{3}{4}b^n$. \\
As a consequence of this and Proposition~\ref{prop:tails},
there exists a constant $K_2$ such
that for each $n$, there is an $\ell$-AUEC code of length
at most $n + \frac{1}{2}\log_{q}n+K_2$. \\
Now let $n$ be a positive integer.
Choose $n_0$ such that
\[ n_0+\frac{1}{2}\log_q n_0+K_2\leq n \mbox{ and }
   (n_0+1) + \frac{1}{2}\log_q(n_0+1) +K_2 \geq n . \]
Our construction shows the existence of an $\ell$-AUEC code of length
$n$ with at least $\frac{3}{4}b^{n_0}$ words.
The definition of $n_0$ implies that
\[ \log_q(n_0+1)\leq\log_q(n+1-\frac{1}{2}\log_q n_0-K_2)
   \leq \log_q(n+1-K_2), \mbox{ and so} \]
\[ n_0 \geq n-1-K_2-\frac{1}{2}\log_q(n_0+1) \geq
               n-1-K_2-\frac{1}{2}\log_q(n+1-K_2) . \]
From the final inequality, it follows that there exists a constant
$K_3$ such that $n_0\geq n-\frac{1}{2}\log_q n - K_3$.
We conclude that
\[\frac{3}{4}b^{n_0}\geq\frac{3}{4}b^n n^{-\frac{1}{2}\log_qb}b^{-K_3}.
  \]
\hfill$\Box$

\section{$\ell$-UEC codes of Varshamov-Tennengolts type}\label{Sec:VT}

In this section we study VT-type $\ell$-UEC codes.
Note however that unlike the VT-codes, the codes
we introduce here are defined by means of some linear equation
(rather than a congruence) over the real field.
Namely given $Q=[0,q-1]\subset\RR$ and $a_0,\dots,a_{n-1},a\in\ZZ$ let
\begin{equation}\label{eq:5.1}
X=\{(x_0,\dots,x_{n-1})\in Q^n:\sum_{i=0}^{n-1}{a_i x_i=a}\}.
\end{equation}

Note that $X$ defines an $\ell$-UEC code if and only if
for each distinct ${\bf x},{\bf y}\in X$ holds
${\bf x}-{\bf y}\notin[-\ell,\ell]^n$
and ${\bf x}-{\bf y}\notin[0,2\ell]^n$.

Thus an obvious sufficient condition for the set of vectors
$X\subset Q^n$
to be an $\ell$-UEC code is that the hyperplane $H$ defined by
$$
H=\left\{(x_0,\dots,x_{n-1})\in\RR^n:\sum_{i=0}^{n-1}a_i x_i=0\right\}$$
does not contain vectors from
$[-\ell,\ell]^n\cup[0,2\ell]^n$, except for the zero vector. \\ \\
An $\ell$-UEC code of VT-type may have the advantage of a simple
encoding and decoding procedure.

In particular, let $\C$ be a code given
by \ref{eq:5.1} where for $i=0,1,\ldots, n-1$, $a_i= (\ell+1)^i$.
In view of observation above  $\C$  is an $\ell$-AEC code.
Suppose now for a received vector ${\bf y}=(y_0,\dots,y_{n-1})$ we have
$$
\sum_{i=0}^{n-1}(\ell +1)^i y_i=a^\prime$$
with $a^\prime\geq a$. Then the transmitted vector
$(x_0,\dots,x_{n-1})=(y_0-e_0,\dots,y_{n-1}-e_{n-1})$,
where the error vector $(e_0,\dots,e_{n-1})$ is just the
$(\ell +1)$-ary representation of the number $a^\prime-a$.

Similarly, if $a^\prime\leq a$, then
$(x_0,\dots,x_{n-1})=(y_0-e_0,\dots,y_{n-1}-e_{n-1})$,
where $(e_0,e_1,\ldots ,e_{n-1})$ is the $(\ell+1)$-ary
representation of $a-a^{\prime}$.

For given $\ell,q$ and $n$, we define
$LA_u(n,\ell)_q=$ the maximum size of an $\ell$-UEC code, over
the alphabet $[0,q-1]$, defined by a linear equation (\ref{eq:5.1}).\\
Correspondingly we use $LA_a(n,\ell)_q$ for $\ell$-AEC codes.

\begin{theo}\label{th:LAu=LAa}
For all $n,q$ and $\ell$, $LA_a(n,\ell)_q=LA_u(n,\ell)_q$.
\end{theo}

{\bf Proof.}
Suppose an $\ell$-AEC code $\C$ is defined by (5.1), that is $\C=X$.
Suppose also w.l.o.g. that $a_0,\dots,a_k<0$ $(k<n-1)$, $a_{k+1},a_{k+1},
\ldots a_n\geq 0$, and $s\triangleq a_0+\dots+a_k$.
Let $\C^\prime$ be the code defined by the equation
\begin{equation}\label{modeq}
-\sum_{i=0}^{k}a_i y_i+\sum_{j=k+1}^{n-1}a_j y_j=a-s(q-1)
\end{equation}

Note that for each ${\bf c}=(c_o,\dots,c_{n-1})\in\C$ the vector
${\bf c}^\prime=(q-1-c_0,\dots,q-1-c_k,c_{k+1},\dots,c_{n-1})\in Q^n$
is a solution of (5.2), that is ${\bf c}^\prime\in\C^\prime$.
The opposite is also true.
Hence we have $|\C|=|\C^\prime|$. Note further that the condition
${\bf c}-{\bf b}\notin[-\ell,\ell]^n$ for each distinct
${\bf c},{\bf b}\in\C$
(this we have since $\C$ is an $\ell$-AEC code) implies that for the
corresponding ${\bf c}^\prime,{\bf b}^\prime\in\C^\prime$ we also have
${\bf c}^\prime-{\bf b}^\prime\notin[-\ell,\ell]^n$. Moreover since
$-a_0,\dots,-a_k,a_{k+1},\dots,a_{n-1}>0$ we have
${\bf c}^\prime-{\bf b}^\prime\notin Q^n$,
which implies that $\C^\prime$ is an $\ell$-UEC code.
Thus we have
$$
LA_a(n,\ell)_q\leq LA_u(n,\ell)_q.$$

This completes the proof since we also have the inverse inequality.
\hfill$\square$ \\ \\
For future reference, we note the obvious fact that for
all $n,\ell,q$ and $q'$, we have
\begin{equation}\label{modeq}
LA_{u}(n,\ell)_q \geq LA_u(n,\ell)_{q'} \mbox{ if } q \geq q'.
\end{equation}

{\bf Remark}
Given $\ell$ and $q$ let $a_0,a_1,\ldots,a_n$ be nonzero
integers such that the code ${\cal C}=X$
defined by (\ref{eq:5.1}) is an $\ell$-UEC code
over the alphabet $Q=[0,q-1]$. Then the following is true.

\begin{prop}\label{prop:Harut}
The code ${\cal C}^*$ defined by
$$
{\cal C}^*=\left\{(z_0,\ldots ,z_{n-1})\in Q^n:
\sum_{i=0}^{n-1}a_iz_i\equiv a\pmod{2\ell S+1}\right\},$$
where $S\triangleq a_0+\cdots +a_{n-1}$ is an $\ell$-UEC code.
\end{prop}

{\bf Proof.}
If for two distinct ${\bf z},{\bf z}'\in{\cal C}^*$ holds
$\sum\limits_{i=0}^{n-1}a_i(z_i-z_i')=0$ then ${\bf z},{\bf z}'$
belong to some translate of code $\C$ and hence
$d_u({\bf z},{\bf z'})\geq2\ell+1$.
Conversely if $\sum\limits_{i=0}^{n-1}a_i({\bf z}-{\bf z'_i})\neq0$
then there exists $j$ (by the pigeonhole principle) such that
$\vert z_j-z_j'\vert\geq 2\ell +1.$
Therefore in both cases $d_u({\bf z},{\bf z}')\geq 2\ell +1.
\hfill\Box$

Thus we have $\vert{\cal C}^*\vert\geq\vert{\cal C}\vert$
which shows that in general the codes given by some congruence
could have better performance.
Note however that by construction given above we cannot have much
gain as compared to the code given by (\ref{eq:5.1}).
This is clear since
$\vert{\cal C}\vert\geq c\vert{\cal C}^*\vert$ for some constant
$c\leq\frac{(q-1)S}{2S\ell +1}<\frac{q-1}{2\ell}.$

\subsection{Lower and upper bounds for $LA_u(n,\ell)_q$}

\begin{theo}\label{LAubound}
For all integers $q,n$ and $\ell$ satisfying $q > \ell + 1$ we have
\[ \frac{\ell}{q-1} \left( \frac{q}{\ell+1}\right)^n
\leq LA_u(n,\ell)_q\leq\lceil\frac q{\ell+1}\rceil^{n-1}. \]
\end{theo}

{\bf Proof.}
Consider the equation
\begin{equation}\label{eq:2.5}
\sum_{i=0}^{n-1}(\ell+1)^i x_i=a,
\end{equation}
and let $X$ be the set of vectors ${\bf x}\in Q^n$
satisfying (\ref{eq:2.5}).
As we have seen in the introduction of this section,
$X$ is a $q$-ary  ${\ell}$-UEC code.

Note also that $X=\varnothing$ if
$a\not\in I\triangleq[0,(q-1)\frac{(\ell+1)^n-1}{\ell}]$.
Hence we infer that there exists an $a\in I$ such that
\[ |X|\geq\frac{|Q^n|}{|I|}=q^n/
\left( (q-1)\frac{(\ell+1)^n-1}{\ell}+1 \right)
\geq \left( \frac{q}{\ell+1}\right)^n \cdot \frac{\ell}{q-1} \; . \]
This gives the lower bound for $LA_u(n,\ell)_q.$\\
Let now $X$ be a $q$-ary $\ell$-UEC code defined by (\ref{eq:5.1}).

To prove the upper bound we consider the mapping
$\psi:Q \to \mathbb{Z} _b$, where
$b\triangleq\lceil\frac{q}{\ell+1}\rceil$, defined by
$$
\psi (j)\equiv j\pmod b; \ j=0,\ldots ,q-1.$$

Correspondingly for a codeword
${\bf x}=(x_0,\ldots ,x_{n-1})\in X$ we define
$\psi ^n({\bf x})=(\psi(x_0),\ldots ,\psi(x_{n-1})$.
Let us show that $\psi ^n$ is an injection on $X$.
Suppose $\psi ^n({\bf x})=\psi ^n({\bf x')}$
for two codewords ${\bf x},{\bf x'}\in X$.
By definition of $\psi $ we have ${\bf x}-{\bf x'}=b{\bf e}$,
where ${\bf e}\in[-\ell,\ell]^n$.
As ${\bf x}$ and ${\bf x'}$ both are in $X$ we have
\begin{equation}\label{eprop}
\sum_{i=0}^{n-1} a_i e_i = 0.
\end{equation}

We define ${\bf x^*}={\bf x'}+(b-1){\bf e}$ and
claim that ${\bf x^*}$ is in $X$.
In view of (\ref{eprop}), it is sufficient to show that
${\bf x^*}\in Q^n$.
For $1\leq i\leq n$ let now $e_i\geq 0$. Then
$x^*_i=x'_i + (b-1)e_i\geq x_i'\geq 0$ and
$x^*_i=x_i-e_i\leq x_i\leq q-1$, so $x^*_i\in Q$.
In a similar way it is proved that $x^*_i\in Q$ if $e_i\leq 0$.
Since ${\bf x}-{\bf x^*}={\bf e}=[-\ell,\ell]^n$, and
{\bf x} and ${\bf x^\ast}$ both are in $X$,
we conclude that {\bf e}={\bf 0}, so ${\bf x}={\bf x'}$.
Thus $\psi ^n$ is an injection, which implies that
$|X|=|\psi ^n(X)|$.

Define now
$$
H'=\{ (y_0,\ldots ,y_{n-1})\in \mathbb{Z} _b^n:
\sum_{i=0}^{n-1} a_iy_i\equiv a(\bmod b)\}.$$

It is easy to see that $\psi ^n(X)\subset H'$.
We can assume without loss of generality that \\
$g.c.d.(a_0,\dots,a_{n-1})=1$, so
$(a_0(\mod b),\dots,a_{n-1}(\mod b))\neq(0,\dots,0)$.

Thus $H'\subset\mathbb{Z}_b^n$ is a
hyperplane over $\mathbb{Z}_b$ and hence
$$
|X|=|\psi ^n (X)|\le
|H'|=b^{n-1}.$$
$\hfill \Box$

\subsection{Construction of optimal codes}

We call a VT-type  $\ell$-UEC code VT-type optimal or shortly
optimal if it attains the upper bound in Theorem~\ref{LAubound}.
In this section we construct, for various classes of pairs
($\ell,q$), maximal $q$-ary $\ell$-UEC codes for each length
$n$. \\
Given integers $\ell\in [1,q-1],\ n,\ r$ we define
\begin{equation}\label{eq:5.8}
{\cal C}_n(r)=\left\{(x_0,\ldots ,x_{n-1})\in Q^n:
\sum_{i=0}^{n-1}(\ell +1)^i x_i=\alpha S_n+r\right\},
\end{equation}
\begin{equation}\label{eq:salphadef}
\mbox{ where }
S_n\triangleq\sum_{i=0}^{n-1}(\ell +1)^i=\frac{(\ell +1)^n-1}{\ell},\
\mbox{ and } \alpha\triangleq\lfloor\frac{q-1}{2}\rfloor.
\end{equation}

As we have seen before,
 \cal ${\cal C}_n(r)$  is an $\ell$-UEC code for all $n$ and $r$. \\
For notational convenience, we denote the cardinality of
 ${\cal C}_n(r)$  by   $\gamma_n(r)$, that is,
\begin{equation}\label{gammadef}
\gamma_n(r) = | {\cal C}_n(r) | \;\;.
\end{equation}

\begin{prop}\label{partition}
For each $n\geq 2$ and each $r$,
\[ \gamma_{n}(r)  = \sum_{x_0}
   \gamma_{n-1}\left((\alpha+r-x_0)/(\ell + 1)\right)  \;\; , \]
where the sum extends over all $x_0\in Q$ satisfying
$x_0\equiv \alpha + r   \pmod{\ell + 1}$.
\end{prop}

{\bf Proof.}
By definition ${\bf x}=(x_0,x_1,\ldots ,x_{n-1})$ is in
${\cal C}_n(r)$ if and only if
$\sum_{i=0}^{n-1} (\ell + 1)^i x_i - \alpha S_n = r$.
Using that $S_n=(\ell + 1)S_{n-1} + 1$, the latter equality
can also be written as
$\sum_{i=1}^{n-1} (\ell + 1)^{i} x_i - \alpha S_{n-1} = r-x_0+\alpha$.
In other words {\bf x} is in ${\cal C}_{n}(r)$ if and only if
$x_0\equiv r+\alpha \;
(\bmod\,\ell + 1)$ and $(x_1,\ldots ,x_{n-1})$ is in
${\cal C}_{n-1}(r')$, where $r'=(r-x_{0}+\alpha)/(\ell + 1)$.
\hfill $\Box$
\\ \\
In the remainder of this section, we use the notation
$\langle x \rangle_{y}$ to denote the integer in $[0,y-1]$
that is equivalent to $x$ modulo $y$. In other words,
$\langle x \rangle_y = x - \lfloor\frac{x}{y}\rfloor \cdot y$.

\begin{lem}\label{Lem1}
Let $e$ and $f$ be integers such that $0\leq e\leq f-1$.
We have that
\[ \vert \{ x\in Q:x\equiv e \;(\bmod f) \}| =
\left\{ \begin{array}{ll}
\lceil\frac{q}{f}\rceil & \mbox{ if } e < \langle q \rangle_{f} \\
\lfloor\frac{q}{f}\rfloor & \mbox{ if } e \geq \langle q \rangle_{f}
\end{array} \right.  \]
\end{lem}

{\bf Proof.}
We obviously have that
\[ \{ x\in Q:x\equiv e \; (\bmod f) \} =
   \{ e+f, e+2f,\ldots ,e+mf\}, \]
where $m$ is such that $e+mf\leq q-1$ and $e+(m+1)f\geq q$.
In other words $m=\lfloor\frac{q-1-e}{f}\rfloor$.
Writing $q=\lambda f + \langle q \rangle_f$, we have
$m - \lambda = \lfloor\frac{\langle q\rangle_f-1-e}{f}\rfloor$,
which equals 0 if $\langle q\rangle_f \geq e+1$, and $-1$ otherwise.
This proves the lemma.
\hfill$\Box$

\begin{theo}\label{extension}
Let $u_1,u_2, \ldots$ and $v_1,v_2,\ldots$ be sequences of integers
such that:
\\ $(1)\;\; 0\leq u_1+\alpha \leq v_1+\alpha\leq q-1$,
\\ and for each $n\geq 2$ \\
$(2)\;\; \lceil\frac{1}{\ell+1}(u_n+\alpha-(q-1))\rceil \geq u_{n-1}$, \\
$(3)\;\; \lfloor\frac{1}{\ell+1}(v_n+\alpha)\rfloor \leq v_{n-1}$,
and  \\
$(4)\;\; \ell+1$ divides $q$,
or for each $r\in [u_n,v_n],\;\;  \langle \alpha+r\rangle_{\ell + 1} <
    \langle q \rangle_{\ell+1}$.

Then for each $n\geq 1$ and $r\in[u_n,v_n]$ we have
$\gamma_{n}(r) = \lceil\frac{q}{\ell+1}\rceil^{n-1}$.
\end{theo}

{\bf Proof.}
We proceed by induction on $n$. \\
For $n=1$ the assertion is true because of condition (1). \\
Now let $n\geq 2$, and suppose the assertion is true for $n-1$.
Let $r\in [u_n,v_n]$. According to Proposition~\ref{partition},
we have that
\begin{equation}\label{part2}
\gamma_n(r) = \sum_{x_0} \gamma_{n-1}
\left( \frac{r+\alpha-x_0}{\ell + 1}\right) \;\; .
\end{equation}

According to condition (4), either $\ell+1$ divides $q$, or
$\langle \alpha+r\rangle_{\ell+1}<\langle q\rangle_{\ell+1}$.
In both cases Lemma~\ref{Lem1} implies that the sum in
(\ref{part2}) has $\lceil\frac{q}{\ell+1}\rceil$ terms. \\
For each $x_0\in Q$ we have that
$r+\alpha-x_0 \leq r+\alpha \leq v_n+\alpha$ and
$r+\alpha-x_0\geq r+\alpha-(q-1)\geq u_n+\alpha-(q-1)$.
That is, for each $x_0\in Q$
\begin{equation}\label{Eq:5.10}
 u_n+\alpha-(q-1) \leq r+\alpha-x_0 \leq v_n+\alpha .
\end{equation}

Combining (\ref{Eq:5.10}) with conditions (2) and (3)
we find that for each $x_0$ in $Q$, such that $r+\alpha-x_0$
is a multiple of $\ell + 1$, we have
\[ \frac{r+\alpha-x_0}{\ell+1} \in [u_{n-1},v_{n-1}] . \]
The induction hypothesis implies that each term in the sum
in (\ref{part2}) equals $\lceil\frac{q}{\ell+1}\rceil^{n-2}$.
\hfill$\Box$

\begin{theo}\label{thm6}
Let $\ell$ and $q$ be such that $\ell+1$ divides $q$.
Let $u_1=-\alpha$, $v_1=\alpha$, and for
$n\geq 2$, $u_n=(\ell+1)u_{n-1}+\alpha$ and
$v_n=(\ell+1)v_{n-1}-\alpha$.
In other words, for
$n\geq 1, v_n=-u_n = \frac{\alpha}{\ell}
\left[ (\ell-1)(\ell+1)^{n-1}+1\right]$.
\\
Then for each $n\geq 1$ and $r\in [u_n,v_n]$, we have
$$
\gamma_n(r) =LA_u(n,\ell )_q=\left(\frac{q}{\ell +1}\right)^{n-1}.$$
\end{theo}

{\bf Proof.}
We apply Theorem~\ref{extension}. It is immediately clear that
conditions (1), (3) and (4) are satisfied. Moreover, for each $n\geq 2$,
$u_n+\alpha-(q-1) = (\ell+1)u_{n-1}+2\alpha-(q-1)\geq
(\ell+1)u_{n-1}-1$, so condition (3) is satisfied as well.
\hfill $\Box$

\begin{theo}\label{Ludores}
Let $c\in[0,\ell], \delta\in \{0,1\}$, and $m$ be such that
\[ q = 2m(\ell+1) + 2c + 1 + \delta \mbox{ and } 2c+\delta \neq \ell . \]
We define $\lambda_1=0$, and for $n\geq 2$,
\[ \lambda_{n} = (\ell+1)\lambda_{n-1} - \eta , \mbox{ where }
   \eta = \left\{ \begin{array}{cl}
    0 & \mbox{ if } 2c+\delta \leq \ell-1, \\
    \lceil\frac{1}{2}(\ell-\delta)\rceil & \mbox{ if } 2c + \delta \geq
            \ell+1 . \end{array} \right. \]
Moreover, for $n\geq 1$, we define
\[ u_n = -c + \lambda_n(\ell+1) \;\mbox{ and } \;
   v_n = -c + \lambda_n(\ell+1)   + \langle q \rangle_{\ell+1}-1 . \]
If $m\leq c-1-\lceil\frac{1}{2}(\ell-\delta)\rceil$ or
$2c+\delta\leq \ell$ and $m\leq c$, then for each integer $n$
and $r\in [u_n,v_n]$,
\[ \gamma_n(r) = LA_u(n,\ell)_q = \lceil\frac{q}{\ell+1}\rceil^{n-1} . \]
\end{theo}

{\bf Proof.}
We apply Theorem~\ref{extension}. Note that
\[ \alpha = \lfloor\frac{q-1}{2}\rfloor = m(\ell+1) + c . \]
We first check condition (1):
$u_1+\alpha = -c + \alpha = m(\ell+1) \geq 0$ and
$u_1+\alpha \leq v_1+\alpha = m(\ell+1) +
\langle q \rangle_{\ell +1} -1 \leq q-1$. \\
The definition of $u_n$ and $v_n$ implies that for each $n$ and
each $r\in [u_n,v_n]$ we have that
\[ r+\alpha \in [u_n+\alpha,v_n+\alpha] =
      [(\lambda_n + m)(\ell+1) , (\lambda_n+m)(\ell+1) +
        \langle q \rangle_{\ell+1} - 1]\; , \]
so condition (4) is satisfied. \\
For verifying Condition (2), we note that
\[ \lceil\frac{1}{\ell+1}(u_n+\alpha-(q-1))\rceil
   = \lceil\frac{1}{\ell+1}(u_n-\alpha-\delta)\rceil =
    (\lambda_n-m) + \lceil\frac{-\delta-2c}{\ell+1}\rceil . \]
As $\lambda_n=\lambda_{n-1}(\ell+1) - \eta = u_{n-1} + c - \eta$
condition (2) is satisfied if and only if
\begin{equation}\label{eq:cond2}
m \leq c - \eta - \lfloor\frac{\delta+2c}{\ell+1}\rfloor.
\end{equation}

For verifying condition (3) we note that
\[ \lfloor\frac{1}{\ell+1}(v_n+\alpha)\rfloor =
   \lfloor\frac{1}{\ell+1}((\lambda_n+m)(\ell+1) +
\langle q \rangle_{\ell+1})\rfloor = \lambda_{n}+m . \]
As $\lambda_n=(\ell+1)\lambda_{n-1} - \eta = v_{n-1}+c -
\langle q \rangle _{\ell+1} + 1 - \eta$,
condition (3) is satisfied if and only if
\begin{equation}\label{eq:cond3}
 m \leq \langle q \rangle_{\ell+1} - 1 -c + \eta
\end{equation}
We distinguish between two cases. \\
{\bf Case 1}
$2c+\delta\leq \ell-1$. \\
Then $\langle q \rangle_{\ell+1}=2c+\delta+1$, and
$\lfloor\frac{\delta+2c}{\ell+1}\rfloor = 0$. That is,
(\ref{eq:cond2}) reduces to the inequality $m\leq c-\eta$ and
(\ref{eq:cond3}) reduces to $m\leq c+\delta+\eta$.
As $\eta=0$, we see that (\ref{eq:cond2}) and (\ref{eq:cond3})
both are satisfied if $m\leq c$. \\
{\bf Case 2}
$2c+\delta\geq \ell+1$. \\
Then $\langle q \rangle_{\ell+1} = 2c+\delta - \ell$, and
$\lfloor\frac{\delta+2c}{\ell+1}\rfloor = 1$. Consequently,
(\ref{eq:cond2}) reduces to the inequality $m\leq c-\eta-1$, and
(\ref{eq:cond3}) reduces to $m\leq c+\delta-\ell-1+\eta$.
With our choice for $\eta$, we see that (\ref{eq:cond2})
and (\ref{eq:cond3}) both are satisfied if
$m\leq c-\eta-1=c-1-\lceil\frac{1}{2}(\ell-\delta)\rceil$.
\hfill $\Box$

\begin{cor}\label{Cor2}
Let $q=(b-1)(\ell+1)+d$ for integers $1\leq b-1<d\leq\ell$. Then for
each $n$
$$
LA_u(n,\ell)_q=b^{n-1}=
\left\lceil\frac q{\ell+1}\right\rceil^{n-1}.$$
\end{cor}

{\bf Proof.}
Suppose $b-1$ is even. Then we can write
$$
q=2m(\ell+1)+d=2m(\ell+1)+2c+1+\delta,$$
where $c=(d-1-\delta)/2$ and $m=(b-1)/2$.
The condition $b-1<d\leq\ell$ implies that $2c+\delta\leq\ell-1$ and
$m\leq c$. Therefore by Theorem 8 we have
$$
\gamma_n(r)=b^{n-1},\text{ where }r\in[-c,c].$$

Suppose now $b-1$ is odd. Then
$$
q=(2m+1)(\ell+1)+d=2m(\ell+1)+d+\ell+1=2m(\ell+1)+2c+1+\delta,$$
where $c=(d+\ell-\delta)/2$ and $m=(b-2)/2$.

Now the condition $b-1<d$ implies
$m\leq c-1-\left\lceil\frac12(\ell-\delta)\right\rceil$
and hence by Theorem 8 we have
$$
\gamma_n(r)=b^{n-1},\text{ where }r\in[u_n,v_n].$$
\hfill$\Box$

In conclusion of this section let us note that the determination of
$LA_u(n,\ell )_q$ in general seems to be a difficult problem.
As was shown above codes defined by (\ref{eq:5.8}) are best possible
for certain parameters $q$ and $\ell$, mentioned in
Theorems 6 and 7.
However we do not know how good these codes are for other parameters.

An interesting open problem is to decide what is the
$\max\limits_r\vert{\cal C}_n(r)\vert$ for given $\ell$ and $q.$ Note
that for some cases the code ${\cal C}_n(0)$ has the size bigger
than the lower bound in Theorem 5. Let for example $\ell =2,\ q=7$.
Then it is not hard to observe that the number of solutions $c_n$ of
(\ref{eq:5.8}) satisfies the recurrence $c_n=2c_{n-1}+c_{n-2}$.
This gives the bound $\vert{\cal C}_n(r)\vert\geq K(2,41)^n,$
where $2,41\approx 1+\sqrt{2}$ is the largest root of the
characteristic equation $x^2-2x-1=0,\ K$ is a constant.
The same recurrence we obtain for any $q=2\ell +3,$ which implies
that for $q=2\ell +3$ and $\ell\geq 2$ one has
$\vert{\cal C}_n(r)\vert\geq K(2,41)^n>
\frac{\ell}{q-1}\left(\frac{q}{\ell +1}\right)^n$
(the lower bound in Theorem 5).
Note however that this is not the case for $\ell =1,\ q=5.$

One can also observe that for $q=7,\ \ell =1$ we have
$\vert{\cal C}_n(r)\vert\geq K(3,51)^n.$
Without going into detail we note that this can be derived from the
recurrence $c_n=4c_{n-1}-2c_{n-2}+c_{n-3}$ for the number of
solutions $c_n$ of (\ref{eq:5.8}) (with $r=0,\ q=7,\ \ell =1$).

One may use a generating functions approach to analize the problem.\\
Let $f(x)$=$1+x+x^2+\ldots +x^{q-1}$.
We are interested in the largest coefficient of the polynomial
$f(x)f(x^{\ell+1})f(x^{(\ell+1)^2})f(x^{(\ell+1)^3}) \cdots
f(x^{(\ell+1)^{n-1}})$. If, for example, we take
$q=5,\ell=1$ and $n=4$,  the largest coefficient equals 20
(attained with $x^{24},x^{28},x^{32}$ and $x^{36}$), while
the coefficient of $x^a$ for
$a=\lfloor\frac{q-1}{2}\rfloor\frac{(\ell+1)^n-1}{\ell}=30$
only equals 17.

\subsection{Asymptotic growth rate of $\ell$-UEC codes of VT type}

In the previous section we explicitly constructed maximal
$q$-ary $\ell$-UEC codes of VT type of arbitrary length
for some classes of pairs ($\ell,q)$ -- but not for all.

In this section we state a less ambitious goal, namely, given
$\ell$ and $q$, to determine the asymptotic behaviour of
$\sqrt[n]{LA_u(n,\ell)_q}$. We will show that this quantity converges
if $n\rightarrow\infty$. As a preparation we need the following

\begin{lem}\label{prop:AB}
Let $a,b,a_0,a_1,\ldots ,a_{m-1},b_0,b_1,\ldots ,b_{n-1}$
be integers such that the codes $\A$ and $\B$, defined as
\[ \A = \{ (x_0,x_1,\ldots ,x_{m-1})\in Q^m:\sum_{i=0}^{m-1} a_ix_i = a\}
\mbox{ and } \B = \{ (y_0,y_1,\ldots ,y_{n-1}) \in Q^n:
\sum_{j=0}^{n-1} b_jy_j = b \} \]
both are non-empty $\ell$-UEC codes.
Let $\A\times \B\subset Q^{m+n}$ be the direct product of $\A$ and $\B$:
\[ \A\times \B =
\{ ({\bf x};{\bf y}):{\bf x}\in \A, {\bf y}\in \B\} . \]
Let $M$ be an integer such that $\sum_{i=0}^{n-1} |a_i|(q-1) < M$, and
define $\C$ as
\[ \C = \{ (z_0,z_1,\ldots ,z_{n+m-1}) \in Q^{n+m}:
\sum_{i=0}^{n-1} a_iz_i \; + \; \sum_{i=n}^{n+m-1}  Mb_{i-n}z_i =
a + Mb \} . \]
Then $\C=\A\times \B$, and $\A\times \B$ is a $q$-ary $\ell$-AUEC code.
\end{lem}

{\bf Proof.}
It is clear that $\A\times \B\subset \C$.
Moreover, $\A\times \B$ is an $\ell$-UEC code:
a received word can be decoded by decoding its $m$ leftmost and
$n$ rightmost symbols to $\A$ and $\B$, respectively.
All we are left with to show is that $\C\subset \A\times \B$.
Therefore, let $(z_0,z_1,\ldots ,z_{n+m-1})$ be in $\C$.
By definition, we have that
\begin{equation}\label{split}
a + Mb = \sum_{i=0}^{m-1} a_iz_i \; + \; M \cdot
\sum_{i=m}^{m+n-1}b_{i-m}z_i,
\end{equation}
and so
\begin{equation} \label{modulo}
a - \sum_{i=0}^{m-1} a_iz_i \equiv 0 \mbox{ mod } M .
\end{equation}

As $\A\neq \emptyset$, there is an ${\bf x}\in Q^m$ such that
$a=\sum_{i=0}^{m-1} a_ix_i$, and whence
\begin{equation} \label{small}
|a - \sum_{i=0}^{m-1} a_iz_i | = | \sum_{i=0}^{m-1} a_i (x_i-z_i) |
\leq \sum_{i=0}^{m-1} |a_i||x_i-z_i| \leq
\sum_{i=0}^{m-1} |a_i|(q-1) < M .
\end{equation}

From (\ref{modulo}) and (\ref{small}) we conclude that
$a = \sum_{i=0}^{m-1}a_iz_i$ and so
$(z_0,z_1,\ldots ,z_{m-1})\in \A$. Furthermore using (\ref{split})
we find that $(z_m,z_{m+1},\ldots ,z_{m+n-1})$ is in $\B$.
\hfill $\Box$ \\ \\
Lemma~\ref{prop:AB} immediately implies that
\begin{equation}\label{supermult}
LA_u(\ell,m+n)_q \geq LA_u(\ell,m)_q \cdot LA_u(\ell,n)_q.
\end{equation}

As $LA_u(\ell,n)_q\leq \lceil\frac{q}{\ell+1}\rceil^{n-1}$
we can invoke Fekete's lemma to derive the following result
from (\ref{supermult}):

\begin{prop}\label{prop:beta}
For each $q$ and $\ell\in Q$, there exists a constant
$\beta(\ell,q)\leq\lceil\frac{q}{\ell+1}\rceil$ such that
\[ \lim_{n\rightarrow\infty} \sqrt[n]{LA_u(n,\ell)_q} =
\beta(\ell,q) . \]
\end{prop}

Theorem~\ref{LAubound} implies that for all $\ell$ and $q$,
\[ \frac{q}{\ell+1} \leq \beta(\ell,q) \leq
\lceil\frac{q}{\ell+1}\rceil . \]
In particular, $\beta(\ell,q)=\frac{q}{\ell+1}$ if
$\ell+1$ divides $q$ (of course, this is also implied by the
much stronger Theorem~\ref{thm6}).
Note also that for pairs $(\ell,q)$ for which the conditions from
Theorem~\ref{Ludores} applies, we have
$\beta(\ell,q)=\lceil\frac{q}{\ell+1}\rceil$.
\\
Inequality (\ref{supermult}) implies that for each $n$,
$\beta(\ell,q)\geq \sqrt[n]{LA_u(n,\ell)_q}$.
For example, consider the case that $q=\ell +2$. The code
\[ \{ (x_0,x_1,x_2,x_3) \in Q^4:\sum_{i=0}^3 (\ell+1)^i x_i =
   \ell+1 + (\ell+1)^3 \} \]
has five words, {\em viz.}
(1+$\ell$,1+$\ell$,$\ell$,0), (1+$\ell$,0,1+$\ell$,0), (1+$\ell$,0,0,1),
(0,1,1+$\ell$,0), and $(0,1,0,1)$.
That is, $\beta(\ell,\ell+2)\geq \sqrt[4]{5}
\approx 1.495$.
Note that Theorem~\ref{LAubound} only allows us to deduce that
$\beta(\ell,\ell+2)\geq \frac{\ell+2}{\ell+1}$. \\
Also note that Corollary~\ref{Cor2} with $b=2$ states that for
$\ell\geq 2$ \ \ $\beta(\ell,\ell+3)=2$.

\section{The error detection problem}\label{Sec:UED}

We find it interesting to consider also the error detection problem,
i.e.\ codes detecting unconventional errors of a certain level.
It is easy to see that codes detecting asymmetric errors of level
$\ell$ can be also used to detect unidirectional errors of level
$\ell$.
For codes detecting all asymmetric (unidirectional) errors of level
$\ell$ we use the abbreviation $\ell$-AED codes
(or $\ell$-UED codes).\\
For integers $\ell,q,n$ satisfying $1\leq\ell<q$ and $n\geq 1$,
we define
$$
P_i=\{(a_1,\dots,a_n)\in Q^n:\sum_{j=1}^n a_j=i\}.$$

It is clear that $P_i$ detect each unidirectional error pattern.
Note that $|P_i|$ is maximal for
$i=i^{\ast}=\lfloor\frac{1}{2}n(q-1)\rfloor$, see
\cite[Thm.\ 4.1.1]{An}. For $a\in [0,\ell n]$, let
$\C_a\subset Q^n$ be defined as
\begin{equation}\label{Cadef}
\C_a= \bigcup_{i:i\equiv a (\bmod\,\ell n +1)} P_i
\end{equation}

\begin{prop}\label{prop:aued}
$\C_a$ is an $\ell$-UED-code over the alphabet $Q$.
\end{prop}

{\bf Proof.}
Clearly $\C_a$ is an $\ell$-UED code iff for each
${\bf x},{\bf y}\in\C_a$ either ${\bf x}$ and ${\bf y}$
are incomparable or $d({\bf x},{\bf y})\geq\ell +1.$
Suppose that for some ${\bf x}=(x_1,\dots,x_n)$ and
${\bf y}=(y_1,\dots,y_n)$ we have ${\bf x}>{\bf y}$.
Then clearly by definition of $\C$ there exists a coordinate
$i\in [1,n]$ such that $x_i-y_i\geq\ell +1$,
i.e.\ $d({\bf x},{\bf y})\geq\ell +1$.
\hfill$\square$

This simple construction gives us a lower bound for the maximum size
of an $\ell$-UED code over alphabet $Q$. However we don't know whether
it is possible to improve this bound, even for the case $\ell=1$.

{\bf Remark 1}.
Asymptotically, taking the union of several $P_i$'s does not really
help as the largest $P_i$ contains $c\frac{1}{\sqrt{n}}q^n$ words,
while nearly all words in $Q^n$ are in the union of about
$\sqrt{n}$ sets $P_i$ with consecutive $i$'s.
\\ \\
{\bf Remark 2}
The construction is not optimal in general. For example take
$\ell$=1 and $q$=$n$=3. It can easily be checked that
($|P_0|,|P_1|,\ldots, |P_6|$) = (1,3,6,7,6,3,1).
Therefore for each $a\in [0,\ell n]=[0,3]$, $|C_a|\leq 7$.
The code consisting of (0,0,0), (2,2,2) and
the six permutations of (0,1,2) has eight words and is a 1-UED code.\\
Consider also two other small cases. \\
For $\ell=1$, $q=4$ and $n=3$ one easily checks that
\\ ($|P_0|,|P_1|,\ldots ,|P_9|$) = (1,3,6,10,12,10,6,3,1) and so
$|C_a|$=16 for all $a\in [0,\ell n] = [0,3]$. \\
Similarly for $\ell$=1, $q$=5 and $n$=3 one easily checks that
\\ ($|P_0|,|P_1|,\ldots, |P_{12}|$) = (1,3,6,10,15,18,19,18,15,10,6,3,1).
It follows that $|C_0|=32$ and $|C_1|=|C_2|=|C_3|= 31$.
Note that $C_0$, the largest of the four codes, does not contain
$P_6$, the largest $P_i$.

\vfill\eject

\end{document}